\documentclass[prl, twocolumn, english, showpacs, floatfix, superscriptaddress]{revtex4}

\usepackage{graphics}
\usepackage{graphicx}
\usepackage{epsfig}
\usepackage{amssymb}
\usepackage{amsmath}

\begin{document}

\newcommand{\ket}[1]{|{#1}\rangle}
\newcommand{\bra}[1]{\langle{#1}|}
\newcommand{\braket}[1]{\langle{#1}\rangle}
\newcommand{\ad}{a^\dagger}
\newcommand{\e}{\ensuremath{\mathrm{e}}}
\newcommand{\norm}[1]{\ensuremath{| #1 |}}
\newcommand{\aver}[1]{\ensuremath{\big<#1 \big>}}
\renewcommand{\Im}{\operatorname{Im}}
\newcommand{\etal}{\textit{et al.}}

%
\title{Getting into Shape: Efficient Cooling Scheme for Fermionic Atoms in Optical Lattices}
\author{Jean-S\'ebastien~Bernier}
\affiliation{Centre de Physique Th\'eorique, CNRS, \'Ecole Polytechnique, 91128 Palaiseau Cedex, France.}
\author{Corinna~Kollath}
\affiliation{Centre de Physique Th\'eorique, CNRS, \'Ecole Polytechnique, 91128 Palaiseau Cedex, France.}
\author{Antoine~Georges}
\affiliation{Centre de Physique Th\'eorique, CNRS, \'Ecole Polytechnique, 91128 Palaiseau Cedex, France.}
\author{Lorenzo~De~Leo}
\affiliation{Centre de Physique Th\'eorique, CNRS, \'Ecole Polytechnique, 91128 Palaiseau Cedex, France.}
\author{Fabrice~Gerbier}
\affiliation{Laboratoire Kastler Brossel, ENS, UPMC, CNRS, 24 rue Lhomond, 75005 Paris, France.}
\author{Christophe~Salomon}
\affiliation{Laboratoire Kastler Brossel, ENS, UPMC, CNRS, 24 rue Lhomond, 75005 Paris, France.}
\author{Michael~K\"ohl}
\affiliation{Cavendish Laboratory, University of Cambridge, Cambridge CB3 0HE, United Kingdom.}

\begin{abstract}
We propose an experimental procedure to cool fermionic atoms loaded into an optical lattice.
The central idea is to spatially divide the system into entropy-rich and -poor regions by
shaping the confining potential profile. Atoms in regions of high entropy per particle are 
subsequently isolated from the system. We discuss how to experimentally carry out this proposal, 
and perform a quantitative study of its efficiency. We find that the entropy per particle, $s$, can typically be reduced 
by a factor of $10$ such that entropies lower than $s/k_B \sim 0.2$ can be reached. Cooling into highly sought-after quantum 
phases (such as an antiferromagnet) can thus be achieved. We show that this procedure is robust against variations of the 
experimental conditions.
\end{abstract}


\maketitle

Rapid experimental progress in manipulating ultracold atomic gases
has provided physicists with increased control over quantum many-particle
systems \cite{BlochZwerger2008}. This was recently evidenced by the observation of a 
Mott-insulating phase of fermionic atoms in a three-dimensional optical 
lattice \cite{JoerdensEsslinger2008,SchneiderRosch2008}. More complex quantum phases, 
such as N\'eel antiferromagnets, strongly correlated Fermi liquids or spin liquids in 
frustrated geometries, could also be realized using cold atoms. However, such phases 
typically emerge in a temperature regime lower than currently achievable. 
In existing experiments, the atomic cloud is pre-cooled by evaporation in a harmonic trap and, in a second step, 
transferred into the periodic potential of an optical lattice. Loading the atoms into
the lattice is ideally performed adiabatically, i.e. conserving the entropy of the system. 
Present experiments indicate an entropy per particle of $s \approx \pi^2 T_o/T_F \approx 1.5-2$  
in the limit of a non-interacting Fermi gas \cite{JoerdensEsslinger2008,SchneiderRosch2008} 
with $T_o$ and $T_{F}$ the system and Fermi temperatures, and $k_B$ set to one \cite{footnote0}. 
These values of $s$ are well above the onset of interesting correlated
phases. Thus, developing novel cooling techniques for lattice quantum gases, as we propose
in this Letter, is a crucial step to demonstrate that cold atoms can indeed adequately
simulate strongly correlated condensed matter systems.

Cooling atomic gases in optical lattices is the focus of an increasing number of studies.
For bosons loaded into an optical lattice, it was proposed to create entropy-rich regions that are 
later isolated from the rest of the system \cite{PoppCirac2006,CapogrossoSvistunov2008}.
These proposals were inspired by earlier experiments in which an adiabatic deformation of the
external trapping potential was used to increase the phase space density of Bose gases
\cite{PinkseWalraven1997,StamperKurnKetterle1998}. For fermions in the absence of 
a lattice potential, it was suggested to cool the gas by taking advantage of a
Feshbach resonance \cite{CarrCastin2004,HaussmannZwerger2008}.
For fermions loaded to an optical lattice very few proposals have been put forward. 
Most of them apply to non-interacting Fermi gases \cite{BlakieBuonsante2007} or are based on the use of a 
Bose-Einstein condensate as a heat reservoir \cite{GriessnerZoller2006,HoZhou2008}.
However, the possible limitation of entropy reduction due to
inelastic collisions between bosons and fermions has not been addressed yet.

In this article, we propose an experimentally realistic procedure to cool two-component fermionic mixtures in optical 
lattices. The key idea is to spatially divide the
trapped fermionic gas into regions of low and high entropy per particle 
by shaping the trapping potential. The two regions are then adiabatically 
isolated from each other and the atoms from the entropy-rich regions are disposed of. 
The remaining atoms have a drastically reduced entropy per particle. In fact, we find 
that the system temperature can be reduced by typically one order of magnitude while retaining half 
of the particles. In addition, the cooling efficiency remains high over a wide range of interatomic 
coupling strengths, initial particle numbers and trap anisotropies.
Hence, with this method, it should be possible to reach highly anticipated 
quantum phases not yet observed. Such phases include the N\'eel antiferromagnet in a cubic 
lattice and, perhaps even more excitingly, spin liquids or other exotic
spin-disordered phases in frustrated lattice geometries \cite{frustratedsyst}. 
Interestingly, 
for systems slightly away from half filling, we can also reach 
sufficiently low entropy per particle to enter the strongly correlated Fermi liquid 
regime. Finally, our proposal, which relies only on adding 
a limited number of lasers to engineer the trap potential, can be well integrated into 
existing experimental setups.

\begin{figure}[tb]
  \centerline{
   \includegraphics[width=1.0\linewidth]{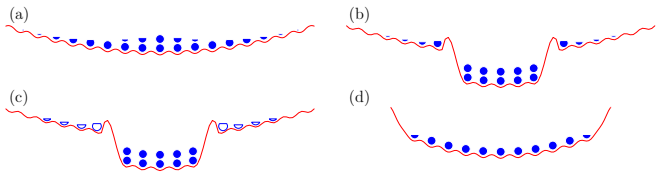}
  }
  \caption{(Color online) Cooling scheme. (a) The atoms trapped in a parabolic profile
are loaded into an optical lattice. (b) A band insulator is created in a dimple at the center
of the trap. This core region is isolated from the rest of the system, the storage region, by rising potential
barriers. (c) The storage region is removed from the system. (d) The band insulator
is relaxed to the desired quantum phase, e.g. a Mott insulator by flattening the dimple and turning off or pushing outwards 
the barriers.}
  \label{fig:scheme}
\end{figure}

\paragraph{Cooling scheme}
Let us begin with a spin-$\frac{1}{2}$ mixture of fermionic atoms pre-cooled in a 
dipole trap.
As a first step, we apply a 
three-dimensional optical lattice potential (Fig.~\ref{fig:scheme} (a)).
To allow the atoms to thermalize, the loading is done in the presence
of a finite but weak interatomic coupling. We also keep the lattice sufficiently
shallow for the atoms to redistribute efficiently. As a second step, we modulate the
entropy distribution by creating a potential depression, a dimple, in the
middle of the harmonic trap. This dimple must be sufficiently deep and narrow for fermionic
atoms to accumulate in it and form a band insulator \cite{footnote1}.
The entropy per particle in this 'core region' is very small.
In contrast, in the outer region, called 'storage region', the potential profile is 
kept shallow in order to create a low density liquid over a wide volume. Under such conditions and at small
interaction strength, the entropy per particle in the storage region is very high.
We then separate the core and storage regions by slowly rising potential barriers, and obtain
the potential profile shown on Fig.~\ref{fig:scheme} (b). As a third step, we 
remove the storage region (Fig.~\ref{fig:scheme} (c)). 
We are left with a new effective system characterized by a very small entropy 
per particle \cite{entropycore}. Finally, as 
a last step, the band insulator is relaxed adiabatically into an experimentally relevant phase. 
For example, if the aim is to form a Mott-insulating state, the filling can be lowered by slowly
flattening the potential in the core region and by turning off or pushing outwards the barriers
(Fig.~\ref{fig:scheme} (d)).

\begin{figure}[tb]
  \centerline{
   \includegraphics[width=1.0\linewidth]{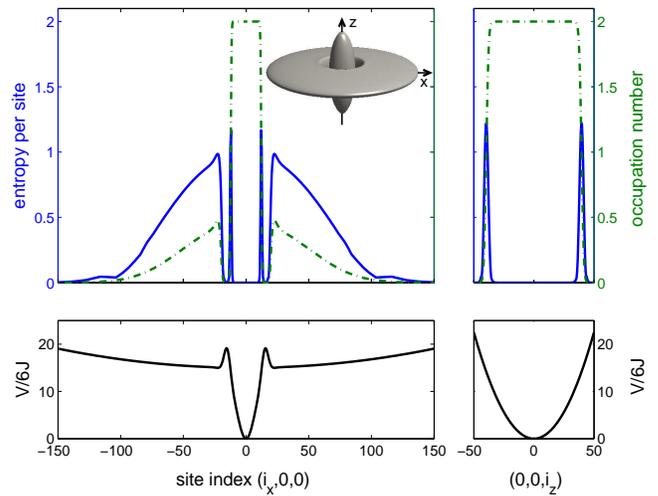}
  }
\caption{
(Color online) Occupation number (dashed line, upper panels), entropy per particle (solid line, upper panels), and
potential profile \cite{footnotefig2} (solid line, lower panels) in the presence of the dimple
and barriers, as a function of the transverse (left) and axial (right) coordinate. 
The potential is offset such that $V=0$ is at the bottom of the dimple. 
We chose the following experimentally realistic parameters: $\frac{U}{6J} = 0.5$, 
$\frac{V_\text{h}}{6J} = 1.8\times 10^{-4}$, $\gamma^2=50$, $\frac{V_{b}}{6J} = 6$,
$r_\text{b} = 15a$, $w_\text{b} = 5a$, $\frac{V_\text{d}}{6J} = 15$, $w_\text{d} = 15a$, and $12 \cdot 10^{4}$ atoms. 
The average entropy per particle in the total system is $s_T = 1.95$ and in the core region $s_C = 0.198$. The ratio of 
particles in the core region versus the total particle number is $\frac{N_C}{N_T} = 0.404$.
Inset: 3d rendering of the potential profile showing an isopotential surface ($\frac{V}{6J}=16$).
}
\label{fig:trap}
\end{figure}

\paragraph{Shaping the potential profile}
The above procedure relies on the ability to add a tailored potential profile on top 
of a lattice potential with amplitude $V_\text{lattice}$. To modulate 
the entropy distribution, the global potential, shown in Fig.~\ref{fig:trap}, should 
realize tight trapping in the core region, surrounded by a wide shallow ring in the storage region isolated 
from the core by high potential barriers. To produce this profile, we envision to use three elements, 
(i) a shallow harmonic trap (either magnetic or optical), (ii) a dimple which confines atoms in a 
small region around the trap symmetry axis and helps to create the band insulator, and (iii) a 
cylindrically-symmetric potential barrier to isolate entropically poor and rich regions. The dimple (ii) and 
potential barrier (iii) are produced by red- and blue-detuned laser beams respectively, creating attractive 
or repulsive dipole potentials. The dimple has a Gaussian profile, while the barrier should rather be 
a narrow annulus. Experimentally this can be realized either by setting a tightly focused laser beam 
in rapid rotation, or by engineering the beam profile using phase plates or other diffractive optics \cite{friedman2002}.
Consequently, in addition to the lattice potential, the trapping profile is given by
\begin{eqnarray}
&&V({\bf r}) = V_\text{harmonic} + V_\text{dimple} + V_\text{barrier} \nonumber
\\
 \textrm{with}
&& V_\text{harmonic}({\bf r}) = V_\text{h}~(x^2+y^2+\gamma^2 z^2)/a^2, \nonumber \\
&& V_\text{dimple}({\bf r}) = -V_\text{d}~\exp{(-2(x^2+y^2)/w_\text{d}^2)}. \nonumber \\
&& V_\text{barrier}({\bf r}) = V_\text{b}~\exp{(-2(\sqrt{x^2+y^2}-r_\text{b})^2/w_\text{b}^2)}, \nonumber
\end{eqnarray}
where $V_{\{\text{h,d,b}\}}$ are the potential amplitudes, $\gamma$ is a measure of the anisotropy
of the harmonic trap, $w_{\{\text{d,b}\}}$ are the waists of the gaussian laser beams forming the dimple and
barrier, $r_b$ is the radius of the cylindrical barrier, and $a$ the lattice spacing.


\paragraph{Efficiency of the procedure}
The efficiency of the proposed cooling scheme can be
quantitatively estimated under the assumption that shaping the
potential profile is an adiabatic process. Possible deviations
from adiabacity will be discussed later on.
Under the adiabatic assumption the
meaningful quantity is the entropy per particle rather than temperature itself.
The cooling efficiency depends on how much entropy per particle is left
in the core region, $s_C=S_C/N_C$, 
at the precise moment when the increasing barrier height causes the two regions to stop exchanging entropy
compared to the initial entropy per particle, $s_T=S_T/N_T$.
The quantities $S_{C/T}$ and $N_{C/T}$ are 
the entropy and number of atoms in the core ($C$) and total system ($T$).
The described situation is shown in Fig.~\ref{fig:trap}.
At later times, the core entropy remains unchanged as the two regions are now isolated
from one another preventing the backflow of entropy.


To determine the efficiency of the cooling scheme, we describe the two-component mixture of fermions 
using a Hubbard-type Hamiltonian \cite{JakschZoller1998}
\begin{eqnarray}
   \label{eq:h}\nonumber
   H =
   -J \sum_{\langle i,j\rangle\sigma} \left(c_{i\sigma}^\dagger
   c^{\phantom{\dagger}}_{j\sigma}+h.c.\right)
   + U \sum _{i} \hat{n}_{i\uparrow} \hat{n}_{i\downarrow} 
   - \sum_{i\sigma}\,\mu_i~\hat{n}_{i\sigma}.
\end{eqnarray}
Here $c^\dagger_{i\sigma}$ and $c_{i\sigma}$ are the creation and annihilation operators of the fermions with $\sigma = \{\uparrow,\downarrow\}$,
$J$ is the hopping matrix element, $U$ is the on-site repulsion, $\mu_i$ is the local
chemical potential and $ \hat{n}_{i\sigma}= c^\dagger_{i\sigma} c^{\phantom{\dagger}}_{i\sigma}$
is the number operator on site $i$. All potential profiles are treated in the local density approximation (LDA),
i.e.~assuming a spatially varying chemical potential $\mu_i =\mu_0 -V_i$. To
use LDA, local densities and entropies must be obtained for the homogeneous system.
These quantities are calculated using dynamical mean
field theory \cite{GeorgesRozenberg1996}. In particular, the entropy is calculated 
as in \cite{DeLeoParcollet2008}.

\begin{figure}[tb]
 \centerline{
 \includegraphics[width=0.8\linewidth]{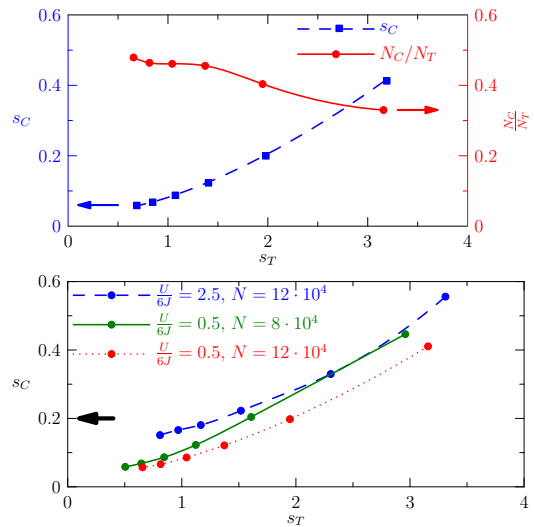}
 }
\caption{
(Color online) Upper panel: entropy per particle in the core, $s_C$, and ratio of particles remaining 
in the core, $N_C/N_T$, after cooling, as a function of the initial total entropy per particle, $s_T$, for 
$\gamma^2=50$, $\frac{U}{6J}=0.5$ and $N_T = 12\cdot10^4$. 
Lower panel: $s_c$ as a function of $s_T$ for different interaction strengths and 
total particle numbers at $\gamma^2=50$. The parameters used to shape the trap are 
the same as in Fig.~\ref{fig:trap} \cite{footnotecaption}. The black arrow in the lower panel 
indicates the pessimistic estimate of the entropy per particle 
required to  reach the antiferromagnet for large interaction strengths \cite{DeLeoParcollet2008}. 
}
\label{fig:gain}
\end{figure}

In the upper panel of Fig.~\ref{fig:gain}, we show, for a three dimensional gas ($\gamma^2=50$), that 
the final entropy per particle in the system core, $s_C$, can be reduced by a factor of ten as compared to 
the initial entropy per particle, $s_T$. This is done while retaining about half of the particles.
For $s_T = 1.95$ and $N_T = 12\cdot10^4$ atoms, about $5\cdot10^4$ atoms are kept. The lower panel of Fig.~\ref{fig:gain} 
shows that the efficiency of our cooling scheme is very stable against variations of the interaction strength 
and initial particle number. Clearly, the procedure is most efficient at small values of the interaction strength, 
but the deviations for other interaction strengths are small. Experimentally a compromise has to be found between a small 
value giving an optimal gain and the time of thermalization for which scattering processes must take place.
Finally, we made sure that this cooling scheme is efficient 
both for quasi two dimensional (large $\gamma^2$) and three dimensional (small $\gamma^2$) systems. However, in two 
dimensions, to obtain similar $N_C$'s and maintain the same efficiency, larger radial 
sizes for both the core and storage regions are required as less particles can be stacked along the $z$ direction. 

The reduction of the entropy per particle by one order of magnitude as 
compared to the current experimental situation opens the door to study a wealth of
unexplored phenomena in cold atomic systems. As an example, in Ref.~\cite{DeLeoParcollet2008},
a (pessimistic) lower bound on the entropy per particle needed to
stabilize antiferromagnetic long-ranged order was estimated to be $s\simeq 0.2$.
Using our cooling scheme, entropies per particle lower than this value can actually
be reached in the core region. Starting from initial temperatures currently accessible 
experimentally, $T_o/T_F \approx 0.15-0.2$ \cite{JoerdensEsslinger2008,SchneiderRosch2008}, 
system temperatures of $T_o/T_F \approx 0.014-0.02$ are achieved.

Finally, two remarks are in order. First, higher efficiencies could be obtained by removing more atoms or engineering 
flatter outer regions that can store more entropy. Second, the weak 
dependence of the efficiency on the initial entropy (Fig.~\ref{fig:gain}) suggests that this procedure 
can as well be performed several times in a row to reach very low temperatures. However, as all changes have to be performed 
slowly, the total time required to cool the system will grow with the number of repetitions.

\paragraph{Removal of storage atoms }
Having shown how the proposed cooling scheme can decrease the entropy in the core region, we now 
adress the fate of the entropy-rich part isolated from the core by the potential barrier. If storage 
atoms do not disturb the subsequent experimental measurements, they can simply be ``pushed outwards'' by dynamically 
increasing the barrier radius and raising its height \cite{footnote3} to avoid ``spilling'' the storage atoms into 
the core region. However, in many cases, getting rid of these atoms or transfering them to a different hyperfine 
state could be advantageous for later detection. Several different removal/transfer schemes may be envisaged 
depending on the details of the experimental setup. These schemes do not need to be adiabatic 
as the storage atoms are already isolated from the core.

One possible removal scheme relies on applying a linear potential gradient 
$-F x$ (which could be due to gravity or to an intentionally applied 
magnetic gradient) and weakening the shallow trap along the $x$-direction. 
Under the influence of the applied force, storage atoms will undergo Bloch oscillations \cite{peik1997} 
interrupted by Landau-Zener (non-adiabatic) transitions to higher bands. 
These transitions can lead to outcoupling of ``atom bursts'' at multiples of the Bloch 
period $T_{B}=h/F a$ \cite{anderson1998}. Atoms in the core region are confined by the 
combined dimple/barrier potentials, and the potential gradient merely shifts the potential 
mininum by a small amount. To achieve significant 
outcoupling rates, one should also significantly lower the lattice depth along $x$. In the weak-binding 
limit, the Landau-Zener formula indeed predicts a transition rate 
$\Gamma_{\rm out} \sim \frac{1}{T_{B}}e^{-A_{LZ}}$ \cite{peik1997}, where $A_{LZ}=m a \Delta^2/4\hbar^2 F$, 
and where $\Delta$ is the bandgap which should be as small as possible. For instance, for a 
lattice depth of $0.5~E_{R}$  ($\Delta\sim 0.2~E_{R}$), $a = 266~$nm and $F/m\sim10~$m/s$^2$, 
we find $\Gamma_{\rm out}\sim 10~$s$^{-1}$ for $^{40}$K atoms ($A_{LZ}\approx 3$) and  essentially 
zero for $^{6}$Li atoms ($A_{LZ}\approx 144$) \cite{footnotebloch}. Another possibility to decrease 
the bandgap is to excite the storage atoms to a higher Bloch band \cite{peik1997}. In order 
to leave the core atoms untouched, the excitation beams should have a ''hollow'' profile (created 
using the same techniques as the potential barriers) to suppress the transition probability 
near the center of the cloud.

\paragraph{Deviation from adiabaticity}
In an experimental setup, one has to find a compromise between changing the potential 
profile slowly (which favors adiabacity) or quickly (which subjects the system to external 
disruptions only for a short time). To approximate the heating induced by non-adiabaticity,
we perform time-dependent simulations of the cooling and subsequent relaxation into a Mott-insulating 
state within an experimentally realistic time. As a measure of the induced heating,
we determine the energy absorbed by the core region during the process.
We use the adaptive time-dependent density matrix renormalization group method \cite{DaleyVidal2004,WhiteFeiguin2004}
to simulate the procedure in a one dimensional fermionic system described by the Hubbard model.
The simulations follow this sequence: we (i) shape the trap (by increasing the dimple amplitude and the
barrier height), (ii) relax the band insulator (by simultaneously pushing outwards the barriers, decreasing
the dimple amplitude, and adjusting the density by changing the parabolic trapping potential), and (iii) tune
the interaction strength to its final value. We assume a linear variation of each parameter with time. 
For a total procedure time of the order of 500 $\frac{\hbar}{J}$ (700 ms for $^{40}$K atoms in a lattice with 
$V_\text{lattice}=8E_R$), the system remains very close to its ground state \cite{footnote6}
and the energy absorbed by the system is smaller by more than one order of magnitude 
than the superexchange coupling $4J^2/U$. Consequently, the heating induced by the non-adiabaticity 
is small enough not to hinder the efficiency of our cooling scheme. We expect that for a three dimensional system the 
timescales for the redistribution of atoms are even more favorable than for the one dimensional case 
simulated here.

\paragraph{Conclusion}
We proposed an efficient scheme to cool fermionic atoms confined to 
optical lattices. This cooling procedure relies on spatially dividing the
trapped fermionic gas into regions of low and high entropy per particle using a complex potential profile. We find
that, for a two-component fermionic mixture loaded into a cubic lattice potential, this scheme 
reduces the system temperature by typically one order of magnitude while keeping approximately half of the atoms. The procedure remains
efficient over a wide range of interatomic coupling strengths, initial particle numbers and
trap anisotropies. This method can be used to cool atoms into highly sought-after quantum phases such as the 
N\'eel antiferromagnet, spin liquids and strongly correlated Fermi liquids.

\paragraph{Acknowledgements}
We acknowledge fruitful discussions with H. Moritz.  Support was provided by
the `Triangle de la Physique', the DARPA-OLE program, ICAM, EPSRC (EP/F016379/1), FQRNT, and 
ANR under contracts GASCOR and FABIOLA.


\begin{thebibliography}{19}
\bibitem{footnotefig2} 
For a 3d lattice loaded with $^{40}$K atoms, we can use beams of $P=2.7$~W at $\lambda_L= 532$~nm focused to a waist of $310~\mu$m. 
For the optical dipole trap, we can use two beams in the $x$ and $y$ directions of $P = 15$~W at  $\lambda_\text{DT} = 1064$~nm 
focused to the same waist as the lattice. $E_R= h^2/(2 m \lambda_L^2)$ is the recoil energy. For $V_\text{lattice} = 8~E_R$, $6J = 0.19~E_R$. 
To create the potential profile, all energy scales were chosen to be smaller than the energy required to excite atoms 
from the first to the second band as $\Delta E_{1 \rightarrow 2}/6J \sim 19$.
\bibitem{footnotecaption} Only the height of the barrier is adjusted to adequately suppress the
entropy exchange between the core and storage regions.
\bibitem{footnote3} For the example shown in Fig.~\ref{fig:trap}, the depth of the shallow trap is on the order 
of $25~J$. A barrier with height exceeding $\gg 36~J$ and radius $\sim 150~a$ implies a realistic tenfold 
increase of laser power compared to the initial configuration.
\bibitem{footnotebloch} The large difference stems from the quadratic dependance of $A_{LZ}$ on the recoil energy, which 
is about 6.7 times larger for $^{6}$Li  than for $^{40}$K.
\bibitem{footnote6} The switching on can be performed on a relatively short time-scale of the order 
of $100~\hbar/J$, whereas the relaxation process is more critical and has been performed on the order of 
$400~\hbar/J$.
\end{thebibliography}

\end{document}